\documentclass[showkeys,showpacs,preprintnumbers]{revtex4}
\usepackage{amsmath,amsfonts,amssymb,amscd,amsxtra,amsthm}
\usepackage{graphicx}
\usepackage{bm}
\usepackage{epstopdf}
\begin{document}
\preprint{KIAS-P11071}
\title{Extended nonlocal chiral-quark model for the $D$- and $B$-meson weak-decay constants}
%-------------------------------------------------
\author{Seung-il Nam}
\email[E-mail: ]{sinam@kias.re.kr}
\affiliation{School of Physics, Korea Institute for Advanced Study (KIAS), Seoul 130-722, Republic of Korea}
%-------------------------------------------------
\date{\today}
\begin{abstract}
In this work, we construct a phenomenological effective model for the heavy-light quark systems, which consist of $(u,d,c,b)$ quarks, i.e. extended nonlocal chiral-quark model (ExNLChQM) to compute the heavy-meson weak-decay constants $f_{D}$ and $f_B$. ExNLChQM is based on the heavy-quark effective field theory as well as the dilute instanton-vacuum configuration. In ExNLChQM, a certain portion of the heavy-meson mass is considered to be generated from the nontrivial QCD vacuum contribution, similar to the light quarks in usual instanton approaches. Hence, the {\it effective} heavy- and light-quark masses become momentum-dependent and play the role of a smooth UV regulator. Employing a generic external-field method applied to the effective action from ExNLChQM, we obtain $f_{D}=(169.28\sim234.57)$ MeV and $f_{B}=(165.41\sim229.21)$ MeV from the numerical results, depending on different model parameters. These values are in relatively good agreement with experimental data and various theoretical estimations. We also discuss the heavy-quark effects on the QCD vacuum, and the decay constants $f_{D^*}$ and $f_{B^*}$ in terms of the heavy-quark spin symmetry. 
\end{abstract}
\maketitle
%--------------------------------------------------
\section{Introduction}
%--------------------------------------------------
For a couple of decades, the heavy-quark effective field theory (HQEFT) has been known to be very powerful theoretical tool to investigate various heavy mesons and baryons, including the charm and bottom quarks~\cite{Georgi:1990um,Neubert:1993mb,Beneke:1994sw}. HQEFT has been well tested in comparison with experiments for those heavy particles by the various  collaborations, such as D$\rlap{/}{0}$~\cite{Abazov:2005ga}, BaBar~\cite{Aubert:2007bx}, Belle~\cite{Ikado:2006un}, BES~\cite{Bai:1999yk}, and so on, and accumulated successful results. HQEFT are based on two heavy-quark symmetries, i.e. spin and heavy-flavor ones. In general, they are not obvious in full QCD and become intact at $m_Q\to\infty$, where $m_Q$ stands for the heavy-quark current mass. As easily understood, the theory contains a small expansion parameter $1/m_Q$ so that QCD can be expanded systematically in terms of it. Additionally, one usually defines the heavy-quark velocity $v$, satisfying $v^2=1$, which designates the velocity superselection rule~\cite{Georgi:1990um}. The heavy-quark field can be redefined with $v$ by integrating out the irrelevant (small) component of the spinor. As a consequence, at the leading $1/m_Q$ expansion, the free heavy-quark Lagrangian manifests the spin and heavy-flavor symmetries. At the hadron level for instance, due to the spin symmetry, there appears mass degeneracy for spin-$0$ and spin-$1$ mesons with the same parity. Interestingly, only a few percent deviation from the mass degeneracy has been observed in the real heavy-meson spectra~\cite{Nakamura:2010zzi}. The heavy-flavor symmetry becomes obvious as well, since the $m_Q$ dependences are removed from the heavy-quark interactions at $v$.

Many effective approaches have been made so far to study the heavy-light quark systems, such as $D$ and $B$ mesons for instance,  based on HQEFT~\cite{Ebert:1994tv,Cvetic:2004qg,Wang:2005qx,Ebert:2006hj,Badalian:2007km,Hwang:2010hw,Penin:2001ux,Geng:2010df}. On top of HQEFT, the light-quark components inside the heavy meson were treated in various ways. Since the light-quarks are believed to be governed dominantly by spontaneous breakdown of chiral symmetry (SBCS), the Nambu-Jona-Lasinio model (NJL) type interactions were employed frequently, i.e. manifesting the chiral symmetry and its dynamical breakdown~\cite{Ebert:1994tv}. Similarly, other nonperturbative QCD approaches were also taken into account to reserch the heavy-light quark systems: Bethe-Salpeter (BS) approach~\cite{Cvetic:2004qg,Wang:2005qx}, relativistic quark model (RQM)~\cite{Ebert:2006hj}, field-correlator method~\cite{Badalian:2007km},  light-cone formalism~\cite{Hwang:2010hw}, QCD sum rule (QCDSR)~\cite{Penin:2001ux}, covariant chiral-perturbation theory (ChPT)~\cite{Geng:2010df}, and so on. As the numerical evaluation of the first principles of QCD, a considerable number of lattice QCD (LQCD) simulations were performed for the related subjects as well~\cite{Becirevic:1998ua,Allton:1990qg,Alexandrou:1990dq,AliKhan:2001jg,Gray:2005ad,Bernard:2009wr}. A mesonic effective interactions were developed for the heavy and light mesons with the flavor symmetry-breaking effects~\cite{Gamermann:2007fi}.

Among the physical observables to be addressed using these effective and LQCD approaches for the heavy-light systems, the heavy-meson weak-decay constants for $D$ and $B$ mesons have been investigated intensively as in the above references, since they are related deeply to the $CP$ violation~\cite{Staric:2011en} as well as the Cabbibo-Kobayashi-Maskawa (CKM) matrix~\cite{Kobayashi:1973fv}. Moreover, the various decay modes for those heavy mesons, such as the (non)leptonic decays, are of importance in understanding the heavy-light systems profoundly. We note that CLEO collaboration have done various experiments for these physical quantities~\cite{:2008sq,Alexander:2009ux,Naik:2009tk}, and the recent experimental data from CLEO were analyzed carefully in Refs.~\cite{Rosner:2008yu,Rosner:2010ak}, giving an estimation $f_{D}=(206.7\pm8.9)$ MeV. Using the $B^-\to\tau^-\bar{\nu}$ data from the Belle~\cite{Ikado:2006un} and BarBar~\cite{Aubert:2007bx} collaborations, the value of  $f_B$ was estimated by $(204\pm31)$ MeV~\cite{Hwang:2010hw}. 

In the present work, considering the increasing interest on the heavy-light quark systems, we want to construct a {\it new} effective model for this subject. We note that the model is based on two ingredients in principle: HQEFT and nonlocal chiral-quark model (NLChQM)~\cite{Praszalowicz:2001wy,Nam:2006sx,Nam:2006mb,Nam:2006au,Dorokhov:2010zz,Dorokhov:2010zzb,Dumm:2010hh,Noguera:2008cm} for the heavy and light quarks, respectively. Since HQEFT has been explained briefly already, we focus on NLChQM here. It is worth mentioning that NLChQM defined in Minkowski space is derived from the instanton-vacuum configuration. Instantons have been believed to be one of the important agents responsible for SBCS in the low-energy region~\cite{Witten:1978bc,Shuryak:1981ff,Diakonov:1983hh}. In the instanton model, the light quark acquires its effective mass $~\sim350$ MeV, which is a consequence of the nontrivial interactions between the quarks and (anti)instantons, as an indication of SBCS as well. Those nontrivial interactions also give the momentum-dependency to the effective-quark mass, which plays the role of a natural UV regulator. There are only two intrinsic parameters in the instanton model: average (anti)instanton size $\bar{\rho}\approx1/3$ fm and inter-(anti)instanton distance $\bar{R}\approx1$ fm~\cite{Shuryak:1981ff}. As long as we are staying at the SU(2) light-flavor sector for $(u,d)$ quarks, these parameters are well determined phenomenologically and can reproduce various nonperturbative-QCD quantities qualitatively very well~\cite{Nam:2006sx,Nam:2006mb,Nam:2006au}. It is worth mentioning that one of the distinctive features in the instanton model is the interactions between the light quarks and pseudoscalar (PS) meson, which takes part in the model by an appropriate SU(2) bosonization process~\cite{Diakonov:1995qy}, becomes nonlocal~\cite{Diakonov:2002fq}. Hence, NLChQM derived from the instanton model bears this distinctive feature naturally. 

Keeping these ingredients in mind, we pay attention to the the mass difference between the heavy meson and heavy quark. In the limit of the infinite heavy-quark current mass, $m_Q\to\infty$, these two masses are the same. However, in reality, one observes they are differ from each other~\cite{Nakamura:2010zzi}. Intuitively, the heavy-meson mass can be regarded as the total sum of the various light and heavy quark masses, i.e. the effective and current ones. Here, we assumed the chiral limit so that the light-quark current mass is neglected. Below is an example for $D$-meson mass:
%EQUATION>>>
\begin{equation}
\label{eq:MASS}
\underbrace{1869\,\mathrm{MeV}}_{D\,\mathrm{meson\,mass}}\hspace{0.7cm}>\underbrace{350\,\mathrm{MeV}}_\mathrm{light\,quark\,effective\,mass}+\underbrace{1290\,\mathrm{MeV}}_\mathrm{heavy\,quark\,current\, mass}.
\end{equation}
%EQUAITON<<<
From Eq.~(\ref{eq:MASS}), we see that about $200$ MeV mass must be generated additionally, even there is a certain value for the binding energy for the heavy meson. In this sense, we assume that this additionally necessary ({\it effective}) mass comes from the similar mechanism with the light quark, i.e. from the nontrivial contributions from the QCD vacuum. If this is the case, we take into account that the effective heavy-quark mass can possess a similar role as a UV regulator as that for the light quark. Considering all these ingredients, we write an effective interactions (action), based on the ideas of HQEFT and NLChQM, extended to the heavy-light quark systems, i.e. extended NLChQM (ExNLChQM).

 In order to test the validity of the present model, we compute the $D$- and $B$-meson weak-decay constants, taking into account their essential implications in physics as mentioned above, using ExNLChQM. We choose all the model parameters from basic NLChQM for the light quarks which are the values for the renormalization scale and light-quark effective-quark mass at $q^2=0$, where $q$ denotes the transferred momentum~\cite{Diakonov:2002fq}. As for the momentum dependences of this mass, we try several possible different choices, as will be discussed in Section II. An unknown parameter from the heavy-quark side, i.e. the effective heavy-quark mass at $q^2=0$ is simply determined to satisfy Eq.~(\ref{eq:MASS}) with small light-quark current mass. As noticed, since the effective masses are decreasing functions and play the roles of UV regulators, there is no need to include any adjustable free parameters by hand, such as the three-dimensional cutoffs. 
 
 Once we obtain the effective action of ExNLChQM, it is an easy task to evaluate the decay constants via conventional functional treatments. From the numerical calculations, we obtain $f_{D}=(169.28\sim234.57)$ MeV and $f_{B}=(165.41\sim229.21)$ MeV, depending on different model parameters for the momentum dependences of the effective heavy and light quark masses as shown in Eq.~(\ref{eq:MMM}) in the next Section. These values are in relatively good agreement with experimental data and various theoretical estimations. By increasing the renormalization scale for the heavy quark, we can obtain more compatible numerical results with the experimental data, which indicates that the heavy quarks make effects on the QCD vacuum to a certain extent, different from the light quark. We also discuss the vector heavy-meson weak-decay constants $f_{D^*}$ and $f_{B^*}$ in terms of the heavy-quark spin symmetry. It turns out that the ratios, $f_{D^*,B^*}/f_{D,B}$ are in $1.01\sim1.03$, which are relatively smaller than those from other theoretical calculations. 

The present work is structured as follows: In Section II, we discuss how to formulate the effective model for the present purpose with brief explanations on HQEFT and NLChQM. Here, we also elucidate the parameters for the present model. Section III is devoted to the numerical results with relevant discussions. We close the present work with summary and future perspectives in the final Section.  

%--------------------------------------------------
\section{General formalism}
%--------------------------------------------------
In this Section, we want to explain the motivation for ExNLChQM and how to construct it based on HQEFT and instanton physics. First, we briefly introduce HQEFT following Refs.~\cite{Georgi:1990um}. The heavy-quark symmetry is fully satisfied in the limit of $m_Q\to\infty$, where $m_Q$ stands for the current-quark mass for the heavy-flavor SU(2) quarks $Q=(c,b)$. Since the top quark, whose mass is about a hundred times larger than these two quarks, we exclude it from the present discussions. Conventionally, the heavy quark filed $Q$ can be rewritten as a {\it soft} one $Q_v$ with a definite velocity $v$~\cite{Georgi:1990um,Neubert:1993mb,Beneke:1994sw}:
%EQUATION>>>
\begin{equation}
\label{eq:HQF}
Q(x)=\frac{1+\rlap{/}{v}}{2}e^{-im_Qv\cdot x}Q_v(x),
\end{equation}
%EQUAITON<<<
where we define the light-like vector $v$ satisfying $v^2=1$. Moreover, we write $v=(1,0,0,0)$ for definiteness, indicating that the heavy quark is at rest. Using this definition of the heavy-quark field, the Dirac equation for it can be modified as~\cite{Georgi:1990um}
%EQUATION>>>
\begin{equation}
\label{eq:DIRAC}
\bar{Q}\left(i\rlap{/}{\partial}-m_Q\right)Q=\bar{Q}_v(iv\cdot\partial)Q_v.
\end{equation}
%EQUAITON<<<
Taking into account the spin symmetry of HQEFT, the pseudoscalar (scalar) and vector (axial-vector) mesons are grouped as a spin doublet  $H$ $(K)$~\cite{Ebert:1994tv}:
%EQUATION>>>
\begin{equation}
\label{eq:HK}
H\equiv\frac{1+\rlap{/}{v}}{2}\left[i\mathcal{P}\gamma_5+\mathcal{V}^\mu\gamma_\mu \right],
\,\,\,\,
K\equiv\frac{1+\rlap{/}{v}}{2}\left[\mathcal{S}+i\mathcal{A}^\mu\gamma_\mu\gamma_5 \right].
\end{equation}
%EQUAITON<<<
Here, $(\mathcal{S},\mathcal{P},\mathcal{V}^\mu,\mathcal{A}^\mu)$ correspond to the fields of the (scalar, pseudoscalar, vector, axial-vector) mesons, consisting of the quark contents $\bar{q}Q$ minimally. In the present work, we will work with the SU(2) light-flavor sector, i.e. $q=(u,d)$. Since we are interested in the weak-decay constants for the PS mesons, such as $D$ and $B$, in order to test the present model, we will concentrate on the doublet field $H$ only hereafter.

Now, we take a look on the light-flavor sector. It has been well known that the light flavors are strongly governed by the spontaneous breakdown of chiral symmetry (SBCS). Among the various chiral effective models, the instanton-inspired models have accumulated successful achievements to investigate important nonperturbative QCD properties and hadrons~\cite{Shuryak:1981ff,Diakonov:1983hh,Diakonov:2002fq}. Based on the dilute-instanton model, the nonlocal chiral quark model (NLChQM) was developed and has been applied to many nonperturbative problems~\cite{Praszalowicz:2001wy,Nam:2006sx,Nam:2006mb,Nam:2006au,Dorokhov:2010zz,Dumm:2010hh,Dorokhov:2010zzb,Dumm:2010hh,Noguera:2008cm}. Considering its successful applications so far, we will employ NLChQM for the light-flavor sectors. The effective chiral Lagrangian of the model has the form as follows:
%EQUATION>>>
\begin{equation}
\label{eq:LFL}
\mathcal{L}^\mathrm{L}_\mathrm{eff}=\bar{q}\left[i\rlap{/}{\partial}-m_q-\sqrt{M_q}^\dagger U_5\sqrt{M_q} \right]q,
\end{equation}
%EQUAITON<<<
where $\hat{m}_q=\mathrm{diag}(m_u,m_d)$ and $M_q$ denote the current-quark and effective masses for the light quarks. According to the isospin symmetry, we choose $m_u=m_d=5$ MeV throughout the present work~\cite{Nakamura:2010zzi}. Note we do not consider the strange quark here whose mass is about $(100\sim200)$ MeV~\cite{Nakamura:2010zzi}, so that the weak-decay constants for $D_s$ and $B_s$ will not be taken into account. The reason for the exclusion of the strange quark is as follows: The inclusion of the strange quark in Eq.~(\ref{eq:LFL}) breaks the validity of the effective Lagrangian in the leading $N_c$ contribution~\cite{Goeke:2007bj}. To remedy this problem, one needs to go over the leading $N_c$ contribution, and it turns out that the meson-loop corrections, corresponding to the large-$N_c$ corrections, are necessary to remedy the problem~\cite{Goeke:2007bj,Nam:2010mh,Nam:2011vn}. Hence, we will confine ourselves to the SU(2) light-flavor sector in the present work to avoid theoretical complexities. Note that $M_q$ is generated from the nontrivial quark-(anti)instanton interactions in terms of the instanton physics, manifesting SBCS  as well as the QCD-vacuum contributions~\cite{Diakonov:2002fq}:
%EQUATION>>>
\begin{equation}
\label{eq:MMMD}
M_q=M_0F^2(t),\,\,\,\,
F(t)=2t\left[I_0(t)K_1(t)-I_1(t)K_0(t)-\frac{1}{t}I_1(t)K_1(t) \right].
\end{equation}
%EQUAITON<<<
We have used the notation $t=|\rlap{/}{\partial}|\bar{\rho}/2$, and $I_\alpha$ and $K_\alpha$ stand for the modified Bessel functions with the order $\alpha$. Note that $U_5$ plays the role of nonlinear PS-meson fields and reads:
%EQUATION>>>
\begin{equation}
\label{eq:U5L}
U_5=\exp\left[\frac{i\gamma_5(\bm{\tau}\cdot\bm{\pi})}{F_\pi} \right]
=1+\frac{i\gamma_5(\bm{\tau}\cdot\bm{\pi})}{F_\pi}-\frac{(\bm{\tau}\cdot\bm{\pi})^2}{2F^2_\pi}\cdots.
\end{equation}
%EQUAITON<<<
Here, $F_\pi$ indicates the light PS-meson weak-decay constant, whose empirical value amounts to $F_\pi\approx92.3$ MeV. As for the interaction Lagrangian in Eq.~(\ref{eq:LFL}), we want to emphasize on that the quark and PS meson interactions are momentum-dependent, i.e. nonlocal interaction, represented by the third term in the square bracket in the right-hand-side of Eq.~(\ref{eq:LFL}).

Taking into account all the ingredients discussed for the light and heavy sectors so far, now, we are in a position to extend this NLChQM to a system with the heavy-light quarks. Our strategy for this purpose is as follows:
%ITEMIZE>>>
\begin{itemize}
\item We extend the flavor SU(2) symmetries for $(u,d)$ and $(c,b)$ into that for SU(4), consisting of $(u,d,c,b)$ quarks.
\item The form of the interaction in Eq.~(\ref{eq:LFL}) is assumed to be kept even for the flavor SU(4), manifesting momentum-dependent quark-PS meson couplings.
\item Hence, the corresponding heavy PS mesons are represented by the nonlinear form as in Eq.~(\ref{eq:U5L}).
\item We consider that the heavy quarks also possess their effective mass as the light one does. We will discuss this in detail soon.  
\end{itemize}
%ITEMIZE>>>
{As has been known well, the flavor SU(4) symmetry is strongly broken in reality, $m_{u,d}\ll m_{c,b}$. For instance, in Ref.~\cite{Gamermann:2007fi}, the mesonic effective Lagrangian was constructed in terms of the SU(4) symmetry with the $(u,d,s,c)$ flavors, and the symmetry-breaking effects were taken into account by suppressing the virtual heavy-meson exchanges $\propto m_{\phi}/m_{\Phi}$, where $m_{\phi}$ and $m_{\Phi}$ denote the light- and heavy-meson masses, respectively. Being different from Ref.~\cite{Gamermann:2007fi}, however, we are treating the quark degrees of freedom (d.o.f.) in the present work, and the symmetry breaking effects are included explicitly and microscopically by the current and effective masses for the heavy and light quarks, although the current heavy-quark mass will be subtracted from the resultant effective Lagrangian by the heavy-quark symmetry~\cite{Georgi:1990um}. Note that the physical heavy-meson mass is used to compute the effective heavy-quark mass as in Eq.~(\ref{eq:HMMM}) as well as the numerical input as in Eq.~(\ref{eq:vp}), manifesting the symmetry-breaking effects. Moreover, the symmetry breaking effects  in the present framework can be deduced from the different UV cutoff masses for the heavy and light sectors, as will be discussed in Section III.}

\begin{figure}[t]
\includegraphics[width=8.5cm]{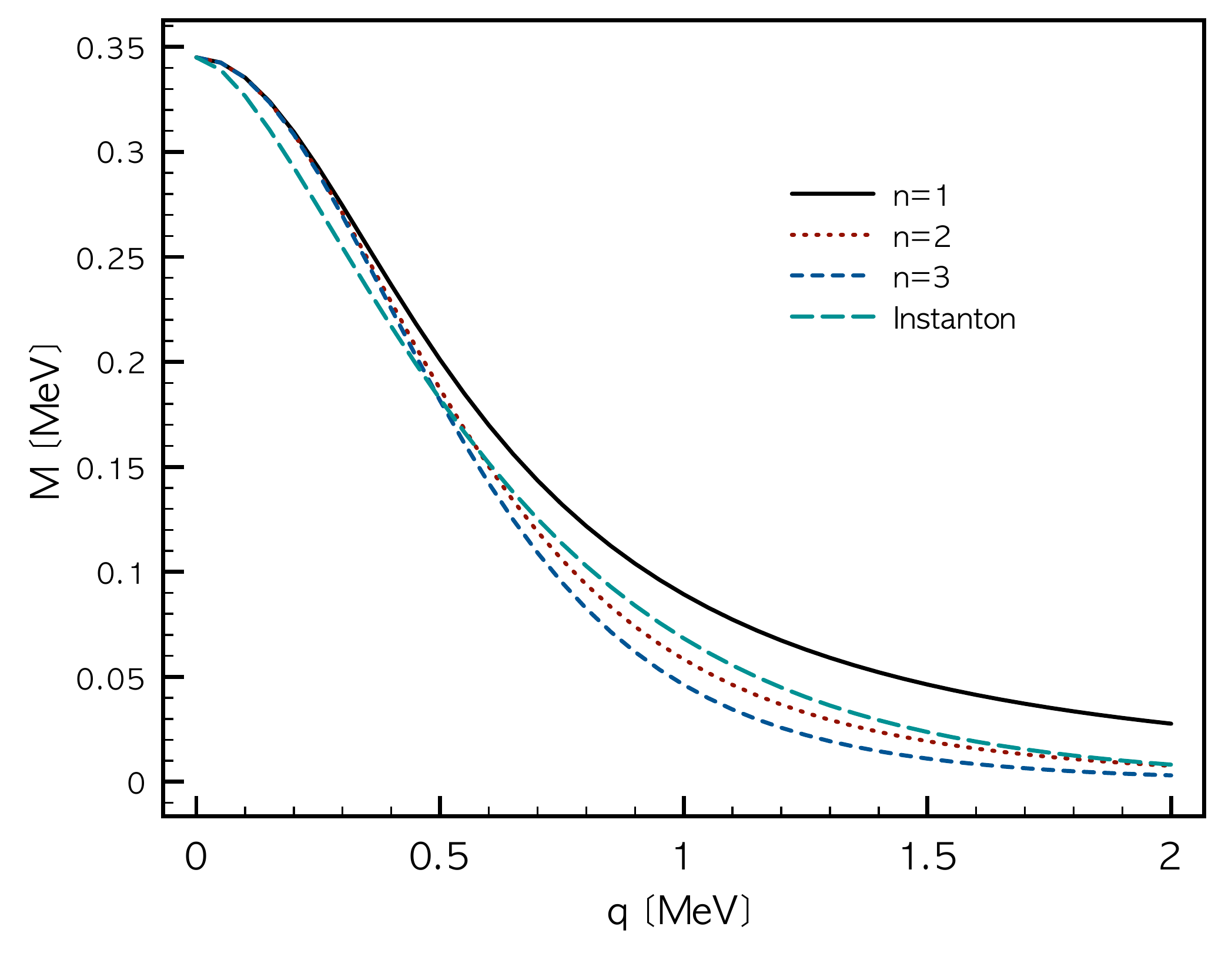}
\caption{Effective light-quark mass $M_q$ as a function of the momentum transfer $q$ in Eq.~(\ref{eq:MMM}) for $n=1$ (solid), $2$ (dot), and $3$ (dash) in Euclidean space. We also draw that from the dilute instanton model (long dash) in Eq.~(\ref{eq:MMMD}). Here, we use $\Lambda=1/\bar{\rho}\approx591$ MeV and $M_{q,0}\approx345$ MeV.}       
\label{FIG2}
\end{figure}

Upon this strategy, we can write an effective Lagrangian density for ExNLCHQM in a neat form as follows:
%EQUATION>>>
\begin{equation}
\label{eq:LAG}
\mathcal{L}^\mathrm{ExNLChQM}_{\mathrm{eff}}=\bar{\psi}
\left[i\rlap{\,/}{D}-\hat{m}_q-\hat{m}_Q-\sqrt{\mathcal{M}}^\dagger\mathcal{U}_5\sqrt{\mathcal{M}} \right]\psi,
\end{equation}
%EQUAITON<<<
where $\psi$ is for the SU(4) flavor spinor, i.e. $\psi=(u,d,c,b)^T$, while $D_\mu$ for the covariant derivative $iD_\mu=i\partial_\mu-e_{(q,Q)}A_\mu$, Here, we use the notations $e_{q,Q}$ for electric charge for a quark $(q,Q)$ and $A_\mu$ for U(1) gauge field. Taking into account the SU(2) isospin and SU(2) flavor symmetries for the light- and heavy-quark sectors~\cite{Georgi:1990um}, we can simply rewrite the spinor for a two-component column matrix as
%EQUATION>>>
\begin{equation}
\label{eq:SPINOR}
\psi=(u,d,b,c)^T\to(q,Q)^T.
\end{equation}
%EQUAITON<<<
The current-quark mass matrices read $\hat{m}_q=\mathrm{diag}(m_u,m_d,0,0)$ and $\hat{m}_Q=\mathrm{diag}(0,0,m_c,m_b)$ again. The square-root of the effective quark-mass matrix $\sqrt{\mathcal{M}}$ is also defined by $\sqrt{\mathcal{M}}=\mathrm{diag}(\sqrt{M_u},\sqrt{M_d},\sqrt{M_c},\sqrt{M_b})$, in which $M_{(q,Q)}$ stand for the effective (light, heavy)-quark mass. In the present work, instead of using Eq.~(\ref{eq:MMMD}), we employ parameterized forms for the effective masses for analytical and numerical convenience as follows:
%EQUATION>>>
\begin{equation}
\label{eq:MMM}
M_{(q,Q)}
=M_{(q,Q),0}\left[\frac{n_{(q,Q)}\Lambda^2}{n_{(q,Q)}\Lambda^2-|i\rlap{/}{\partial}|^2} \right]^{n_{(q,Q)}},
\end{equation}
%EQUAITON<<<
where $M_Q$ corresponds to the soft-component of the heavy-quark field $Q_v$ as understood. Note that, from now on, we choose $n_{q,Q}$ as positive-integer free parameters. As a trial, we choose the cases for $(n_q,n_Q)=1$ and $(n_q,n_Q)=2$, being assigned as {\it Model I} and {\it Model II}, respectively. As mentioned in the previous Section, $\Lambda$ denotes the renormalization scale of the present model and relates to the inverse of the average (anti)instanton size $\bar{\rho}\approx1/3$ fm, resulting in $\Lambda\approx591$ MeV. In Figure~\ref{FIG2}, we depict the effective light-quark mass in Eq.~(\ref{eq:MMM}) as a function of the momentum transfer $q$ for $n=1$ (solid), $2$ (dot), and $3$ (dash) for $M_{q,0}=345$ MeV in Euclidean space ~\cite{Nam:2006au}. We also draw that from the dilute-instanton model (long dash), given in Eq.~(\ref{eq:MMMD}). As shown in the figure, when $n=2$, the parameterized mass is very similar to that from the instanton model. Moreover, as $n$ increases, the mass curves get decreasing faster. As {\it Model III}, we set $n_q=2$ for the light quark and $n_Q=1$ for the heavy quark, i.e. $(n_q,n_Q)=(2,1)$, considering a possibility that the momentum dependences for the heavy and light quarks may different from each other. We will not take into account the case with $(n_q,n_Q)=(1,2)$, since the numerical results are almost the same with those from Model III. The reason for this similarity can be easily understood by seeing Eqs.~(\ref{eq:WDC}) and (\ref{eq:FEX}) in Section III, since the term $(M_qM_Q)^{1/2}$ in the numerator governs the difference between Models.
%FIGURE>>>
%FIGURE<<<

The nonlinear PS-meson field $\mathcal{U}_5$ in a $(4\times4)$ matrix form reads:
%EQUATION>>>
\begin{equation}
\label{eq:U5}
\mathcal{U}_5=
\left(
\begin{array}{cc}
\exp\left[\frac{i\gamma_5\bm{\tau}\cdot\bm{\pi}}{F_{\pi}} \right]
&\exp\left[\frac{i\gamma_5\bm{\tau}\cdot\bm{\Pi}}{F_{\Pi}} \right]-{\bf 1}_{2\times2}\\
\exp\left[\frac{i\gamma_5\bm{\tau}\cdot\bar{\bm{\Pi}}}{F_{\Pi}} \right]-{\bf 1}_{2\times2}
&\exp\left[\frac{i\gamma_5\bm{\tau}\cdot\bm{\Pi'}}{F_{\Pi'}} \right]\\
\end{array}
\right)
\to\left(
\begin{array}{cc}
{\bf 1}_{2\times2}+\frac{i\gamma_5\bm{\tau}\cdot\bm{\pi}}{F_\pi}
&\frac{i\gamma_5\bm{\tau}\cdot\bm{\Pi}}{F_\Pi}\\
\frac{i\gamma_5\bm{\tau}\cdot\bar{\bm{\Pi}}}{F_\Pi}
&{\bf 1}_{2\times2}\\
\end{array}
\right),
\end{equation}
%EQUAITON<<<
where $F_{\pi,\Pi,\Pi'}$ indicate the weak-decay constants for the isovector PS mesons $(\pi,\Pi,\Pi')$, consisting of $(\bar{q}q, \bar{q}Q\,\mathrm{or}\,\bar{Q}q, \bar{Q}Q)$ quark contents. Note that we have off-diagonal terms in $\mathcal{U}_5$ which are necessary for the heavy-light quark mixing terms. The subtraction in the off-diagonal terms, i.e. $-{\bm1_{2\times2}}$, indicates that there is no flavor mixing between heavy and light quarks. We expand $\mathcal{U}_5$ up to $\mathcal{O}(\pi,\Pi,\Pi')$, since we will compute the matrix element $\langle0|J_W|\mathrm{PS\,meson}\rangle$ for $F_{D,B}$. Moreover, because we are not in interested in the quarkonia states such as $J/\Psi\sim c\bar{c}$ and $\Upsilon\sim b\bar{b}$, which are about twice-times heavier than $D$ and $B$ mesons, we set the lower-right $(2\times2)$ block matrix to be a unit matrix, resulting in a Lagrangian for the heavy quarks in Eq.~(\ref{eq:HQF}). However, it is a straightforward task to include those quarkonia states in the present framework. The explicit expressions for the $(2\times2)$ block flavor matrices in $\mathcal{U}_5$ are written with the light and heavy PS mesons as follows:
%EQUATION>>>
\begin{equation}
\label{eq:TAU}
\bm{\tau}\cdot\bm{\pi}=\left(
\begin{array}{cc}
\pi^0&\sqrt{2}\pi^+\\
\sqrt{2}\pi^-&-\pi^0
\end{array}
\right),\,\,\,\,
\bm{\tau}\cdot\bm{\Pi}=\sqrt{2}\left(
\begin{array}{cc}
\bar{D}^0&B^+\\
D^-&B^0
\end{array}
\right),\,\,\,\,
\bm{\tau}\cdot\bar{\bm{\Pi}}=\sqrt{2}\left(
\begin{array}{cc}
D^0&D^+\\
B^-&\bar{B}^0
\end{array}
\right).
\end{equation}
%EQUAITON<<<
Also, the explicit expression for $\mathcal{U}_5$ is given in Appendix in a $(4\times4)$ matrix form. {Note that, if we replace the present $(u,d,c,b)$ flavors into $(u,d,s,c)$, one recovers the SU(4) PS-meson fields given in Ref.~\cite{Gamermann:2007fi}. It is worth mentioning that Eq.~(\ref{eq:LAG}) breaks chiral symmetry explicitly, whereas it is dynamically broken by the finite $M_{q,Q}$ values, which indicate the nontrivial QCD vacuum effects~\cite{Shuryak:1981ff,Diakonov:1983hh,Diakonov:2002fq}. This chiral-symmetry breaking pattern is inherited from NLChQM beyond the chiral limit~\cite{Nam:2006au}.} 

Using the definitions from Eq.~(\ref{eq:SPINOR}) to Eq.~(\ref{eq:TAU}), the effective Lagrangian in Eq.~(\ref{eq:LAG}) can be represented in three separate parts, i.e. light-light (LL), heavy-heavy (HH), and heavy-light (HL, LH) quark terms:
%EQUATION>>>
\begin{eqnarray}
\label{eq:EFL1}
\mathcal{L}_\mathrm{eff}^\mathrm{ExNLChQM}
&=&\mathcal{L}^\mathrm{LL}_\mathrm{eff}
+\mathcal{L}^\mathrm{HH}_\mathrm{eff}+\mathcal{L}^\mathrm{(HL,LH)}_\mathrm{eff}
\cr
&=&
\left[\bar{q}\left(i\rlap{/}{\partial}_q-m_q-M_q \right)q
-\frac{1}{F_\pi}\bar{q}\sqrt{M_q}
\left[\gamma_5(i\bm{\tau}\cdot\bm{\pi}) \right]
\sqrt{M_q}q \right]_\mathrm{LL}
+\left[\bar{Q}\left(i\rlap{/}{\partial}_Q-m_Q-M_Q \right)Q\right]_\mathrm{HH}
\cr
&-&\left[\frac{1}{F_\Pi}\bar{Q}\sqrt{M_Q}
\left[\gamma_5(i\bm{\tau}\cdot\bm{\Pi})
 \right]
\sqrt{M_q}q\right]_\mathrm{HL}
-\left[\frac{1}{F_\Pi}\bar{q}\sqrt{M_q}
\left[(i\bm{\tau}\cdot\bar{\bm{\Pi}})\gamma_5 \right]
\sqrt{M_Q}Q\right]_\mathrm{LH},
\end{eqnarray}
%EQUAITON<<<
where all the masses in Eq.~(\ref{eq:EFL1}) are $2\times2$ diagonal matrices for each flavor SU(2). Being similar to the heavy-quark field in Eq.~(\ref{eq:HQF}), the {\it zero-spin} heavy PS-meson field can be redefined with its soft component as follows:
%EQUATION>>>
\begin{equation}
\label{eq:HMFD}
\bm{\Pi}=e^{-im_Qv\cdot x}\bm{\Pi}_v,\,\,\,\,
\bar{\bm{\Pi}}=e^{im_Qv\cdot x}\bar{\bm{\Pi}}_v.
\end{equation}
%EQUAITON<<<
It is an easy task to redefine $\mathcal{L}^\mathrm{HH}_\mathrm{eff}$ in Eq.~(\ref{eq:EFL1}) using the Dirac equation for the heavy quarks in Eq.~(\ref{eq:DIRAC}), resulting in
%EQUATION>>>
\begin{equation}
\label{eq:HHL}
\mathcal{L}^{\mathrm{HH}}_\mathrm{eff}=\bar{Q}_v\left[\rlap{/}{v}(iv\cdot\partial)-M_Q\right]Q_v=\bar{Q}_v\left[(iv\cdot\partial)-M_Q\right]Q_v.
\end{equation}
%EQUAITON<<<
In the second step of Eq.~(\ref{eq:HHL}), we use the projection relation $\rlap{/}{v}Q_v=Q_v$~\cite{Georgi:1990um}. Using Eqs.~(\ref{eq:HQF}) and (\ref{eq:HMFD}), we can rewrite $\mathcal{L}^{\mathrm{(HL,LH)}}_\mathrm{eff}$ in the following form:
%EQUATION>>>
\begin{equation}
\label{eq:HLLH}
\mathcal{L}^{\mathrm{HL}}_\mathrm{eff}=-\frac{1}{F_\Pi}
\bar{Q}_v\sqrt{M_Q}
\left[\frac{1+\rlap{/}{v}}{2}
\gamma_5(i\bm{\tau}\cdot\bm{\Pi}_v) \right]
\sqrt{M_q}q,\,\,\,\,
\mathcal{L}^{\mathrm{LH}}_\mathrm{eff}=-
\frac{1}{F_\Pi}\bar{q}\sqrt{M_q}
\left[(i\bm{\tau}\cdot\bar{\bm{\Pi}}_v)\frac{1+\rlap{/}{v}}{2}\gamma_5 \right]
\sqrt{M_Q}Q_v,
\end{equation}
%EQUAITON<<<
which satisfies $\mathcal{L}^{\dagger\mathrm{(HL,LH)}}_\mathrm{eff}=\mathcal{L}^{\mathrm{(LH,HL)}}_\mathrm{eff}$. According to the spin symmetry for the heavy quarks and comparing with Eq.~(\ref{eq:HK}), the heavy PS meson can be re-defined in a isotriplet spin-doublet meson for $(0^-,1^-)$, as follows:
%EQUATION>>>
\begin{equation}
\label{eq:DOU}
\frac{1+\rlap{/}{v}}{2}\gamma_5(i\bm{\tau}\cdot\bm{\Pi}_{v})\to
\frac{1+\rlap{/}{v}}{2}\gamma_5
[i\bm{\tau}\cdot\bm{\mathcal{P}}+\gamma_\mu(\bm{\tau}\cdot\bm{\mathcal{V}}^\mu)]
\equiv \bm{\tau}\cdot\bm{H}.
\end{equation}
%EQUAITON<<<
Using a generic functional integral technique for the Grassmann variables given in Appendix, finally, we can arrive at an effective action for the heavy-light quark systems from the effective Lagrangian density in Eq.~(\ref{eq:EFL1}):
%EQUATION>>>
\begin{eqnarray}
\label{eq:EFA3}
&&\mathcal{S}^\mathrm{LL+HL+LH}_\mathrm{eff}=
\cr
&&-i\mathrm{Sp}\ln
\Bigg[i\rlap{/}{\partial}-\bar{M}_q-\frac{1}{F_\pi}\sqrt{M_q}(i\gamma_5\pi)
\sqrt{M_q}-\left(\frac{1}{F_H}\sqrt{M_Q}H\sqrt{M_q}
\right) (iv\cdot\partial-M_Q)^{-1}
\left(\frac{1}{F_H}\sqrt{M_q}\bar{H}\sqrt{M_Q}\right)\Bigg],
\end{eqnarray}
%EQUAITON<<<
where $\bar{M}_q=m_q+M_q$, and $F_H$ denotes the weak-decay constant for the spin-doublet mesons as in Eq.~(\ref{eq:DOU}). We have used notations $\pi\equiv(\bm{\tau}\cdot\bm{\pi})$ and $H=\bm{\tau}\cdot\bm{H}$ for convenience. It is worth noting that the effective action in Eq.~(\ref{eq:EFA3}) is in principle equivalent to the first term of Eq.~(36) in Ref.~\cite{Ebert:1994tv}, except for the momentum dependent quark-PS meson coupling strengths. 

As a next step, we estimate $M_{Q,0}$ in Eq.~(\ref{eq:MMM}) from a simple phenomenological analysis. In this consideration, the heavy PS-meson mass can be understood as
%EQUATION>>>
\begin{equation}
\label{eq:HMMM}
M_H\approx
\left[m_q+M_{q,0} \right]_\mathrm{L}+\left[m_Q+M_{Q,0} \right]_\mathrm{H},
\end{equation}
%EQUAITON<<<
where we have ignored the binding energy for the meson. From the experimental data for $D$ and $B$ mesons, we can write
 %EQUATION>>>
\begin{eqnarray}
\label{eq:ES}
M_D&=&1869.57\,\mathrm{MeV}\approx(m_c+m_q+M_{q,0}+M_{Q,0})=1625.0\,\mathrm{MeV}+M_{Q,0}\to M_{Q,0}\approx 244.57\,\mathrm{MeV}
\cr
M_{B}&=&5279.17\,\mathrm{MeV}\approx(m_b+m_q+M_{q,0}+M_{Q,0})=5025.0\,\mathrm{MeV}+M_{Q,0}\to M_{Q,0}\approx 254.17\,\mathrm{MeV}.
\end{eqnarray}
%EQUAITON<<<
The numerical inputs, which are taken from Ref.~\cite{Nakamura:2010zzi}, are summarized in Table~\ref{TABLE1}. As shown in Eq.~(\ref{eq:ES}), it is necessary to add the {\it effective} heavy-quark mass to reproduce the heavy-meson mass appropriately. To estimate the values of $M_{Q,0}$ above, we choose $M_{q,0}\approx350$ MeV as a trial. If we think the binding energy for the mesons, the estimated value $M_{Q,0}=(240\sim250)$ MeV must be the lower bound of its real one. From these observations and previous discussions, we assume the following two:
%ITEMIZ>>>
\begin{itemize}
\item  The finite effective heavy-quark mass, $M_{Q,0}$ is generated from the same mechanism with that for the light quark-instanton interaction, representing the nontrivial QCD vacuum effects. For the numerical calculations, $M_{Q,0}$ is estimated by Eq.~(\ref{eq:HMMM}).
\item The instanton ensemble is not affected much by the heavy sources such as the heavy quarks, resulting in that one can use $\Lambda\approx1/\bar{\rho}\approx600$ MeV again for Eq.~(\ref{eq:MMM}). However, we will also discuss the possibility for $\Lambda$ for the heavy quark to be changed from about $600$ MeV in Section III. 
\end{itemize}
%ITEMIZ>>>
Hence, we use the effective heavy-quark mass in the momentum space as in Eq.~(\ref{eq:MMM}), although we do not go over its microscopic derivation, which must be beyond the scope of the present work. 

%TABLE>>>
\begin{table}[h]
\begin{tabular}{c|c||c|c}
$m_c$&$M_D$&$m_b$&$M_B$\\
\hline
$(1290.0^{+0.05}_{-0.11})$ MeV&$(1869.57\pm0.16)$ MeV&$(4670.0^{+0.18}_{-0.06})$ MeV&$(5279.17\pm0.29)$ MeV\\
\end{tabular}
\label{TABLE1}
\caption{Numerical input values for the heavy quarks and mesons for $F_{D,B}$ in Eq.~(\ref{eq:WDC}), taken from Ref.~\cite{Nakamura:2010zzi}. Here, we choose the masses for the charged heavy mesons.}
\end{table}
%TABLE>>>

%--------------------------------------------------
\section{Numerical Results and discussions}
%--------------------------------------------------
%FIGURE>>>
\begin{figure}[t]
\includegraphics[width=5cm]{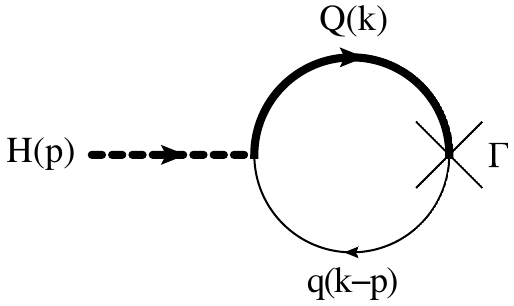}
\caption{Feynman diagram for the weak decay for a heavy meson $H=(D,B)$ with the momentum $p$. Thick and thin solid lines indicate the heavy $Q=(c,b)$ and light $q=(u,d)$ quarks, respectively. $\Gamma$ denotes an arbitrary Lorentz vertex depending on the decay processes.}       
\label{FIG1}
\end{figure}
%FIGURE<<<
In this Section, we present numerical results, computed using the effective action of ExNLChQM in Eq.~(\ref{eq:EFA3}), to verify the validity of the present model. For this purpose, we perform simple calculations for the heavy PS-meson decay constants $F_{D,B}$ through a conventional functional method. First, we define the $(B,D)\equiv H$ meson weak-decay constants as follows~\cite{Ebert:1994tv}:
%EQUATION>>>
\begin{equation}
\label{eq:FPI}
\langle0|\bar{q}(x)\gamma_\mu(1-\gamma_5)\tau^aQ_v(x)|H^b(p)\rangle
=2ip_\mu F_H\delta^{ab},
\end{equation}
%EQUAITON<<<
where the superscripts $(a,b)$ indicate the isospin indices, while $p$ is the on-mass shell momentum of $H$ with the velocity $v$. To evaluate the matrix elements in Eq.~(\ref{eq:FPI}), it is convenient to employ the external-current method, which can be represented by the following effective action: 
%EQUATION>>>
\begin{eqnarray}
\label{eq:EFA4}
\mathcal{S}^\mathrm{LL+HL+LH}_\mathrm{eff}&=&
-i\mathrm{Sp}\ln
\Bigg[i\rlap{\,/}{D}-\bar{M}_q-\frac{1}{F_\pi}\sqrt{M_q}(i\gamma_5\pi)
\sqrt{M_q}+\alpha_\pi\cdot J^a_\pi
\cr
&&-\left(\frac{1}{F_H}\sqrt{M_Q}H\sqrt{M_q}
+\alpha_H\cdot J^a_H\right) (iv\cdot\partial-M_Q)^{-1}
\left(\frac{1}{F_H}\sqrt{M_q}\bar{H}\sqrt{M_Q}
+\alpha_{\bar{H}}\cdot J^a_{\bar{H}}\right)\Bigg].
\end{eqnarray}
%EQUAITON<<<
Here, $\alpha$ and $J^a$ denote the relevant external source and current with a certain isospin index $a$, respectively, in general. Although we do not evaluate $F_\pi$ numerically using the effective action in Eq.~(\ref{eq:EFA4}) in the present work, since we have done it several times in the previous works~\cite{Nam:2008xx}, the numerical result for $F_\pi$ turned out to be about $93$ MeV, which is in a very good agreement with its empirical value, with the phenomenological instanton parameters $\Lambda=591$ MeV and $M_{q,0}=345$ MeV. These values will be employed further for $F_{D,B}$.

To evaluate $F_{D,B}$, one needs to determine the value for $v\cdot p$, considering the definitions in Eq.~(\ref{eq:FPI}). From the on-mass shell condition for the meson $H$ at $v$~\cite{Ebert:1994tv},  one is lead to 
%EQUATION>>>
\begin{equation}
\label{eq:vp}
v\cdot p\approx\Delta M_{D,B}=M_{D,B}-m_{c,b}.
\end{equation}
%EQUAITON<<<
We note that this quantity in Eq.~(\ref{eq:vp}) describes the long-distance {\it nonperturbative} effects~\cite{Penin:2001ux}. Hence, taking the heavy-quark rest frame $v_\mu=(1,0,0,0)$, one can write $p_\mu=(\Delta M_H,0,0,0)$. Performing standard functional derivatives with respect to $\alpha_H$ and $\bar{H}$ for the weak current in Eq.~(\ref{eq:FPI}), the analytic expression for $F_H$ can be obtained as follows:
%EQUATION>>>
\begin{equation}
\label{eq:WDC}
F^2_H=\frac{4N_c}{v\cdot p}\int\frac{d^4k}{(2\pi)^4}
\frac{\sqrt{M_{q,k-p}M_{Q,k}}}
{[(k-p)^2-\bar{M}^2_{q,k-p}]}
\frac{v\cdot k-v\cdot p}{v\cdot k-M_{Q,0}+i\epsilon}.
\end{equation}
%EQUAITON<<<
Note that, in the right-hand-side of Eq.~(\ref{eq:WDC}), we replace the effective heavy-quark mass into that at $k^2=0$ in the denominator, i.e. $M_{Q}\to M_{Q,0}$, since this replacement does not make any considerable changes, while the numerical calculations get easier to a great extent. It is worth mentioning that, due to the decreasing behavior of the effective quark masses in Eq.~(\ref{eq:MMM}) with respect to momentum transfer, the integration over $k$ in Eq.~(\ref{eq:WDC}) does not diverge even without any cutoffs included by hand. The integration in Eq.~(\ref{eq:WDC}) can be evaluated by 1) performing the Wick rotation to Euclidean space and 2) Cauchy integral over $k'_4\equiv k_4-M_{Q,0}$, as done in Ref.~\cite{Ebert:1994tv}. After these analytic calculations, we arrive at an expression for the heavy-meson weak-decay constant:
%EQUATION>>>
\begin{eqnarray}
\label{eq:FEX}
F^2_H&=&\frac{N_c}{\Delta M_H}\int\frac{\bm{k}^2d\bm{k}}{\pi^2}
\frac{\sqrt{M_{q,0}M_{Q,0}}(\Delta M_H-M_{Q,0})
[\bm{k}^2+(M_{Q,0}-\Delta M_H)^2+n_q\Lambda^2]^{2n_q}}
{[\bm{k}^2+(M_{Q,0}-\Delta M_H)^2+n_q\Lambda^2]^{2n_q}[\bm{k}^2 +(M_{Q,0}-\Delta M_H)^2]+(n_q\Lambda^2)^{2n_Q}\bar{M}^2_{q,0}}.
\cr
&\times&\left[\frac{(n_q\Lambda^2)^{n_q/2}}
{[\bm{k}^2+(M_{Q,0}-\Delta M_H)^2+n_q\Lambda^2]^{n_q/2}}
 \right]\left[\frac{(n_Q\Lambda^2)^{n_Q/2}}{(\bm{k}^2+n_Q\Lambda^2)^{n_Q/2}},
 \right],
\end{eqnarray}
%EQUAITON<<<
where $n_{q,Q}$ are defined with Eq.~(\ref{eq:MMM}).

%FIGURE>>>
\begin{figure}[t]
\begin{tabular}{cc}
\includegraphics[width=8.5cm]{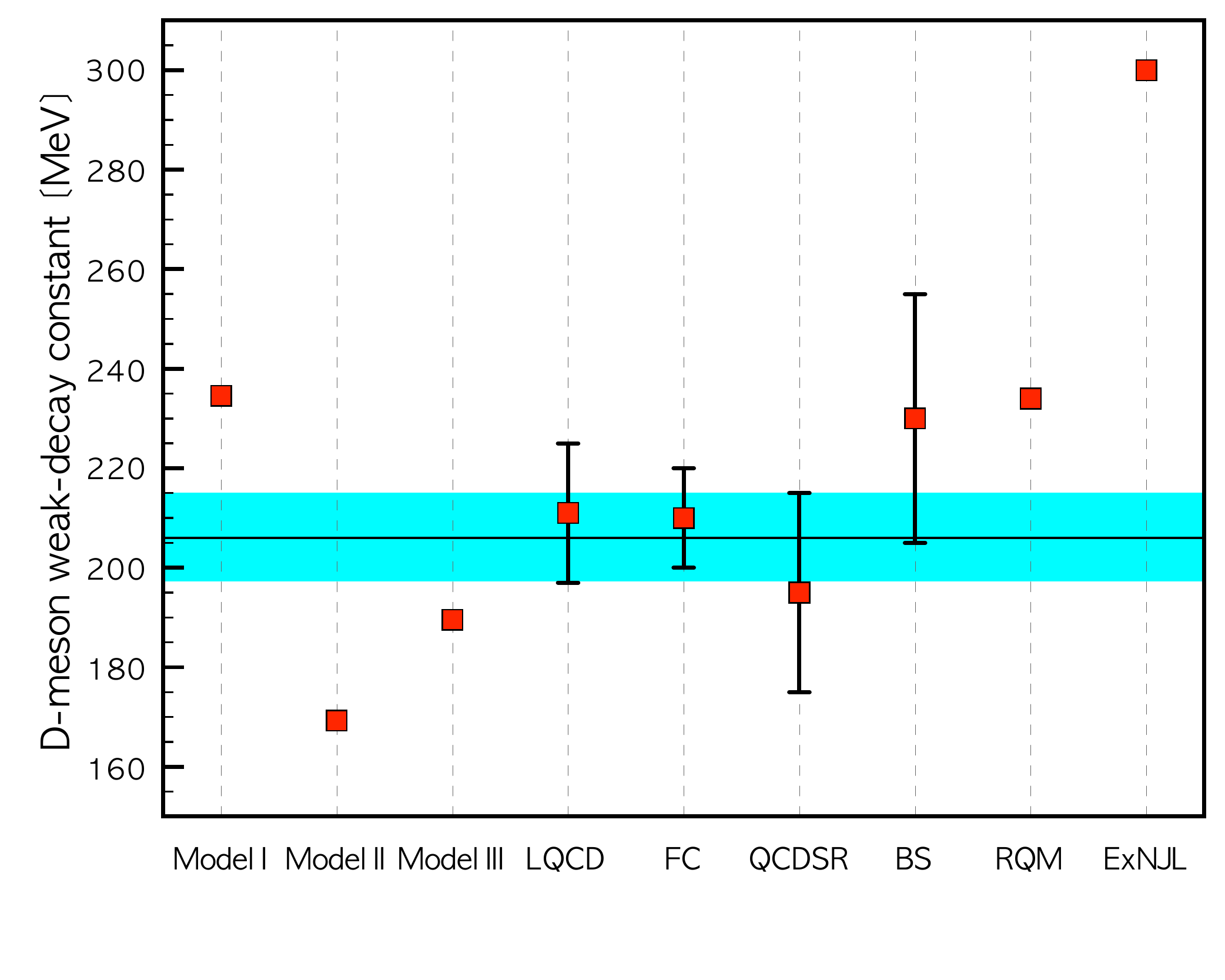}
\includegraphics[width=8.5cm]{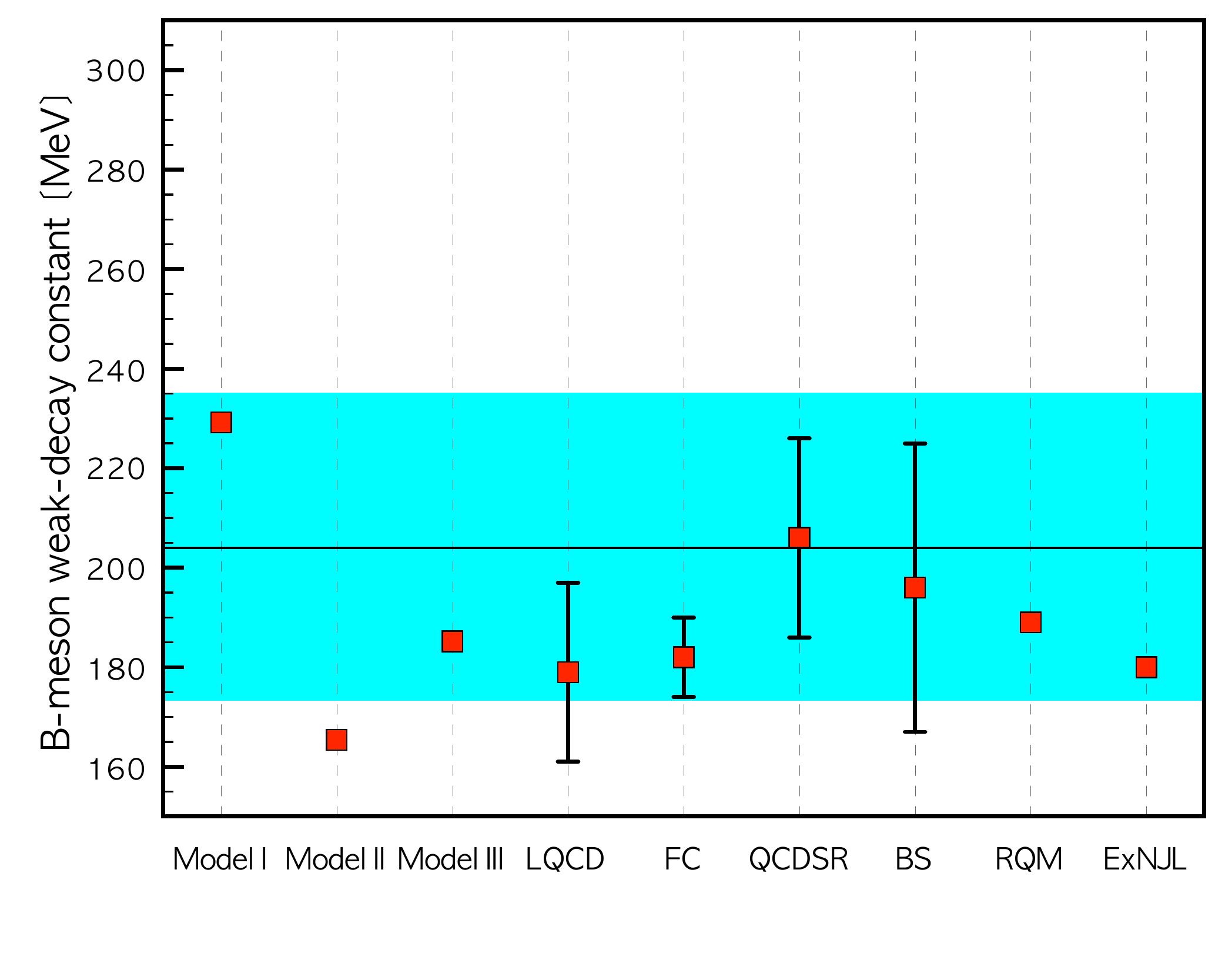}
\end{tabular}
\caption{$f_D$ (left) and $f_B$ (right) computed by various theoretical approaches, which are indicated on $x$ axis and given in Table~\ref{TABLE2}. The shaded areas stand for the experimental data with corresponding errors~\cite{Hwang:2010hw,Rosner:2008yu,Rosner:2010ak}. The horizontal thin-solid lines indicate the center value of the experimental data.}        
\label{FIG34}
\vspace{0.5cm}
\begin{tabular}{cc}
\includegraphics[width=8.5cm]{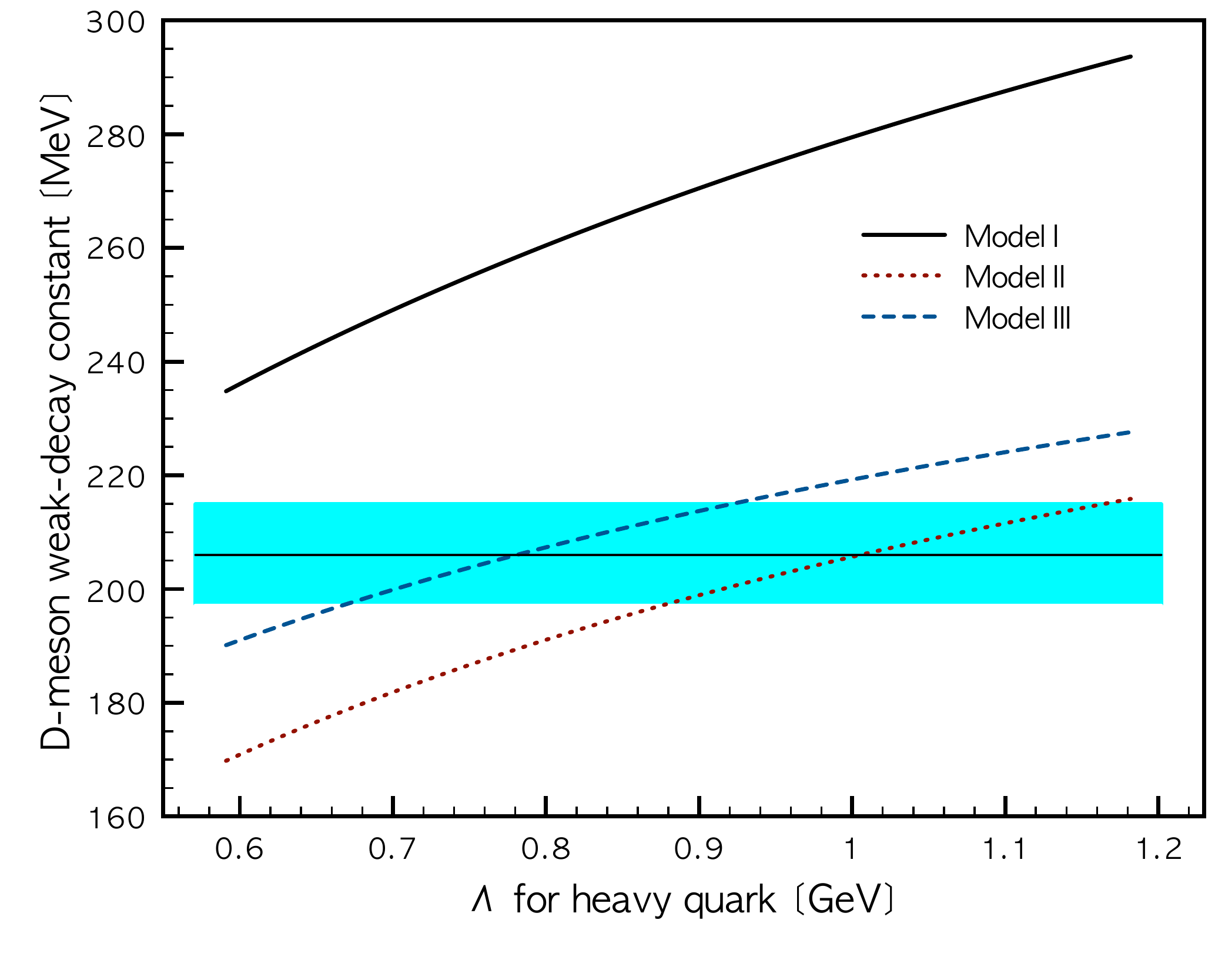}
\includegraphics[width=8.5cm]{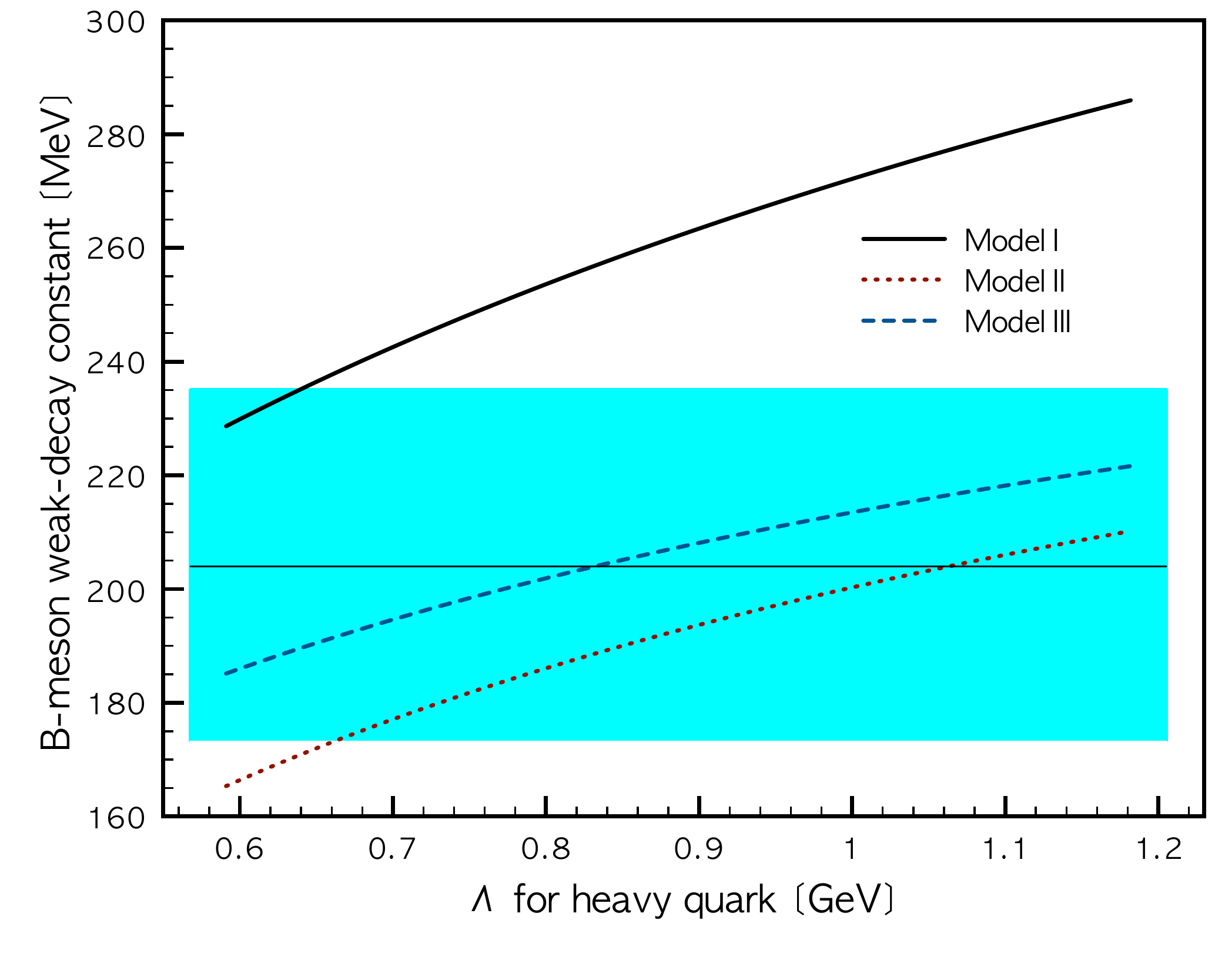}
\end{tabular}
\caption{$f_D$ (left) and $f_B$ (right) for different renormalization scale $\Lambda$ in Eq.~(\ref{eq:MMM}) for the heavy quark. Here, we fix it for the light quark as $\Lambda=591\approx1/\bar{\rho}$ MeV. The shaded areas stand for the experimental data with corresponding errors~\cite{Hwang:2010hw,Rosner:2008yu,Rosner:2010ak}. The horizontal thin-solid lines indicate the center value of the experimental data.}       
\label{FIG56}
\end{figure}
%FIGURE<<<

In order to compare the present numerical results with other theoretical ones, we normalize them alternatively as $F_H\to f_H\equiv\sqrt{2}F_H$. Then we have the following numbers for the heavy-meson weak decay constants, $f_{D,B}$, for Model I with $(n_q,n_Q)=1$:
%EQUATION>>>
\begin{equation}
\label{eq:MODEL1}
f_D= 234.57\,\mathrm{MeV},\,\,\,\,\,\,\,\,f_B=229.21\,\mathrm{MeV}.
\end{equation}
%EQUAITON<<<
Similarly, those values computed via Model II with $(n_q,n_Q)=2$ read as follows:
%EQUATION>>>
\begin{equation}
\label{eq:MODEL2}
f_D= 169.28\,\mathrm{MeV},\,\,\,\,\,\,\,\,f_B=  165.41\,\mathrm{MeV},
\end{equation}
%EQUAITON<<<
and, from Model III with $(n_q,n_Q)=(2,1)$, we have
%EQUATION>>>
\begin{equation}
\label{eq:MODEL3}
f_D= 189.53\,\mathrm{MeV},\,\,\,\,\,\,\,\,f_B=  185.20\,\mathrm{MeV}.
\end{equation}
%EQUAITON<<<
Note that all the numerical results of Model I, II, and III are evaluated at $\Lambda=591$ MeV for the both heavy and light quarks. If we take into account the errors in heavy-quark and heavy-meson masses~\cite{Nakamura:2010zzi}, these values would be changed by $\pm(2\sim3)$ MeV. In Table~\ref{TABLE2}, we list experimental and other theoretical results in comparison with ours. For this table, we refer mainly to Ref.~\cite{Hwang:2010hw}. From the table we see that our Model-I results are in about $10\%$ deviations from the experimental data $f_{(D,B)}=(206\pm8.9,204\pm31)$ MeV~\cite{Rosner:2008yu,Rosner:2010ak}, whereas those from Model II considerably underestimates the experimental data by about $20\%$. Model III results locate in the middle of those from Model I and II, as expected, and closer to the experimental data than others. The theoretical results from the lattice QCD (LQCD) with the improved clover action in the quenched approximation, they had $f_{(D,B)}=(211\pm14,179.0\pm18)$ MeV~\cite{Becirevic:1998ua}. A field-correlator method (FC) was also employed then gave $f_{(D,B)}=(210\pm10,182\pm8)$ MeV~\cite{Badalian:2007km}. Comparing with these values, our results are relatively compatible with them, showing only several percent differences. The QCD sum rule (QCDSR), $f_{(D,B)}=(195\pm20,206\pm15)$ MeV~\cite{Penin:2001ux}, and Bethe-Salpeter method (BS), $f_{(D,B)}=(230\pm25,196\pm29)$ MeV~\cite{Cvetic:2004qg,Wang:2005qx}, also present the comparable results with ours. Note that there is tendency $f_D<f_B$ from the QCDSR results, being different from others. The relativistic quark model (RQM)~\cite{Ebert:2006hj} and extended NJL model (ExNJL)~\cite{Ebert:1994tv} present relatively large values for $f_D$ in comparison to other estimations. Note that the theoretical framework of Ref.~\cite{Ebert:1994tv} is in principle the same with ours, except for the quark-meson coupling schemes. Although we did not list in Table~\ref{TABLE2}, there are many other LQCD results, such as $f_B=(204\pm8)$ MeV~\cite{AliKhan:2001jg} and $(216\pm9)$ MeV~\cite{Gray:2005ad}, which locate inside the experimental errors. 

In order to test the $\Lambda$ dependence for the heavy quark, considering the possibility for the QCD vacuum to be affected by the heavy sources differently from the light quarks, in Figure~\ref{FIG56}, we plot the curves for $f_D$ (left) and $f_B$ (right) as functions of the renormalization scale $\Lambda$ in Eq.~(\ref{eq:MMM}) for the heavy quark. Here, the renormalization scale for the light quark remains unchanged, i.e. $\Lambda\approx591$ MeV, since it has been well known from the usual instanton models theoretically as well as phenomenologically~\cite{Shuryak:1981ff,Diakonov:1983hh,Diakonov:2002fq}. In general, as $\Lambda$ increases, the values for $f_{D,B}$ also does smoothly. As shown in the figure, Model I results are almost excluded from the experimental data for any of $\Lambda$ values. On the contrary, if $\Lambda$ becomes about $1.0\,(0.8)$ GeV for Model II (III), one obtains considerably good agreement with the data. From this observation, one can argue that the renormalization scale for the heavy quark can be modified by the interaction between the heavy quarks and QCD vacuum into a larger one, i.e. $\Lambda\sim0.6\,\mathrm{GeV}\to1\,\mathrm{GeV}$, which may indicate the effects of the heavy sources to the vacuum. Since $\Lambda$ is proportional to the inverse size of (anti)instanton, the increasing of $\Lambda$ stands for the spatial shrinkage of the (anti)instantons. We notice that this tendency is consistent with that the effective heavy-quark mass $\sim250$ MeV, estimated by Eq.~(\ref{eq:ES}), is smaller than the effective light-quark mass $\sim350$ MeV, since the absolute value for the effective mass is proportional to the interaction strength between the quarks and (anti)instantons. As understood by seeing Figure~\ref{FIG56}, if we tune the $\Lambda$ values for the heavy quark, we can reproduce $f_{D,B}$ being compatible very well with the center values of the data. For instance, we obtain $f_{(D,B)}=(205.00,200.32)$ MeV with $\Lambda=1.0$ GeV for the heavy quark via Model II. Note that these values show only a few percent deviations from the experimental data. However, we do not provide those numerical results in the present work, because we are interested in showing the validity of ExNLChQM, not in those fine tunings. We want to leave such a phenomenological studies for the heavy-light systems for the future works.

By construction of the present model with the spin symmetry, the decay constants for the vector heavy mesons $f_{D^*}$ and $f_{B^*}$ can be computed using Eq.~(\ref{eq:FEX}) as well. Only difference between the pseudoscalar and vector mesons in the present theoretical framework is their masses as the inputs. Here are some discussions for the vector mesons: We note that the vector-meson d.o.f. are not incorporated with generic NLChQM, since all other fields except for the PS one are integrated out~\cite{Goeke:2007bj}, resulting in that the present ExNLChQM is not accompanied with them by construction. There have been several attempts to take into account the vector mesons within NLChQM in terms of the vector-meson dominance (VMD)~\cite{Dorokhov:2004ze}. In that way, ExNLChQM can be modified for $D^*$ and $B^*$ without considering the spin doublets in HQEFT. One can also include the heavy vector-meson fields nonlinearly in the present model as in Eq.~(\ref{eq:U5}), being similar to Ref.~\cite{Gamermann:2007fi}. In order to test the VMD-modified ExNLChQM, one may compute $f_{D^*,B^*}$ as well as the heavy vector-meson light-cone wave functions~\cite{Hwang:2010hw}. We would like to leave this VMD modification of ExNLChQM as future works. 

As the numerical results, ignoring the errors and uncertainties in the quark mass values, we have $f_{D^*}=(240.72,\,172.74,\,194.52)$ MeV and $f_{D^*}=(231.80,\,167.31,\,187.31)$ MeV for Model (I, II, III). These computed values for the vector mesons are also listed in Table~\ref{TABLE2} with comparison with other theoretical estimations. In Ref.~\cite{Hwang:2010hw}, the values for $f_{(D,B)^*}$ were computed by using the experimental values for $f_{(D,B)}$, which are underlined in the table, as inputs. Note that, from our model,  the values for $f_{(D,B)^*}$ are about a few percent larger than those for the PS mesons. Similar tendency, $f_{(D,B)^*}>f_{(D,B)}$, is also observed in other theoretical calculations as in the table. The ratio for the PS and vector heavy-meson decay constants, $f_{(D,B)^*}/f_{(D,B)}$ is a good quantity to see this tendency quantitatively. As seen in the table, other theoretical calculations provide the ratio, which ranges from $1.1$ to $1.48$ with about $10\%$ uncertainties. Being different from them, our numerical results show small numbers for the ratios as explained above. This smallness may indicate again the necessity to modify $\Lambda$ for the heavy quark.

From these observations discussed above, we can conclude that the present model, i.e. ExNLChQM reproduces the weak-decay constants $f_{D,B}$ qualitatively, in comparison to various experimental and theoretical results. Moreover, we have shown that there is a more room to improve the model from a phenomenological point of view by considering the heavy-quark effects on the nontrivial vacuum. 
 %TABLE>>>
\begin{table}[h]
\begin{tabular}{c||c|c|c||c|c|c|c|c|c|c}
[MeV]&Model I&Model II&Model III&LCF~\cite{Hwang:2010hw}&LQCD~\cite{Becirevic:1998ua}&FC~\cite{Badalian:2007km}&QCDSR~\cite{Penin:2001ux}&BS~\cite{Cvetic:2004qg,Wang:2005qx}&RQM~\cite{Ebert:2006hj}&ExNJL~\cite{Ebert:1994tv}\\
\hline
\hline
$f_D$&$234.57$&$169.28$&$189.53$
&$\underline{206\pm8.9}$~\cite{Rosner:2008yu}&$211\pm14$&$210\pm10$
&$195\pm20$&$230\pm25$&234&300\\
$f_B$&$229.21$&$165.41$&$185.20$&$
\underline{204\pm31}$~\cite{Rosner:2010ak}&$179\pm18$&$182\pm8$
&$206\pm20$&$196\pm29$&189&180\\
\hline
$f_{D^*}$&$240.72$&$172.74$
&$194.52$
&$259.6\pm14.6$&$245\pm20$&$273\pm13$
&$\cdots$&$340\pm23$&268&$\cdots$\\
$f_{B^*}$&$231.80$&$167.31$&
$187.31$&$225\pm38$&$196\pm24$&$200\pm10$
&$\cdots$&$238\pm18$&219&$\cdots$\\
\hline
\hline
$f_{D^*}/f_D$&$1.03$&$1.02$&$1.03$&$1.26\pm0.12$&$1.16\pm0.14$&$1.30\pm0.13$&$\cdots$&$1.48\pm0.09$&$1.15$&$\cdots$\\
\hline
$f_{B^*}/f_B$&$1.01$&$1.01$&$1.01$&$1.10\pm0.16$&$1.09\pm0.13$&$1.10\pm0.13$&$\cdots$&$1.21\pm0.06$&$1.16$&$\cdots$\\
\end{tabular}
\caption{The values for $f_{D,B}$ and $f_{D^*,B^*}$ from experimental~\cite{Rosner:2008yu,Rosner:2010ak} (underlined and inputs for Ref.~\cite{Hwang:2010hw}) and theoretical results [MeV]. We also summarize the ratios $f_{D^*,B^*}/f_{D,B}$ for various cases. All the numerical results of Model I, II, and III are evaluated at $\Lambda=0.591$ MeV for the both heavy and light quarks. The explanations for the abbreviations and numerics are given in the text.}
\label{TABLE2}
\end{table}
%TABLE>>>

%--------------------------------------------------
\section{Summary and outlook}
%--------------------------------------------------
In this work, we have constructed a phenomenological heavy-light quark effective model, i.e. ExNLChQM and tested it to reproduce the heavy-meson weak-decay constants. The present model was based on and motivated by the heavy-quark effective field theory and instanton-vacuum configuration. In what follows, we list some important points of ExNLChQM and observations of the present work:
%TABLE>>>
\begin{itemize}
\item We focused on the SU(4) heavy-light flavor sector, i.e. $(u,d,c,b)$ quarks, since the inclusion of the strange quark breaks down the validity of the usual leading-$N_c$ instanton-induced effective model and the top quark is too heavy to be incorporated with these four quarks. The flavor SU(4) symmetry-breaking effects are taken into account explicitly by the current and effective masses for the heavy and light quarks, using the physical heavy-meson masses as numerical inputs. 
\item We assumed that the heavy quark also possesses its effective mass, generated from the interactions with the nontrivial QCD vacuum, being similar to the light quarks, resulting in the momentum-dependent effective heavy-quark mass, which plays the role of a UV regulator. As a consequence in the present model, the heavy meson and heavy-light quarks are coupled nonlocally. 
\item The renormalization scale of the model was set to be $\Lambda\approx600$ MeV, which relates to the inverse of the average (anti)instanton size. It was also approximated that the $\Lambda$ for the heavy-quark effective mass is the same with that for the light quark, taking into account that the QCD-vacuum structure is not affected much by the heavy quarks, although the heavy sources may distort the vacuum to a certain extent. 
\item On top of the heavy-quark spin symmetry for the heavy mesons, the $D$- and $B$-meson weak-decay constants are computed using the ExNLChQM effective action, resulting in $f_{D}=(169.28\sim234.57)$ MeV and $f_{B}=(165.41\sim229.21)$ MeV from the numerical results, depending on the instanton parameters. These values are compatible with experimental and other theoretical estimations. 
\item As the renormalization scale $\Lambda$ for the heavy quark gets increased, one obtains more compatible results with the data, although we do not provide those numerical numbers. This tendency indicates that the QCD vacuum can be affected by the heavy sources with different strengths in comparison to the light-quark case. 
\item The vector heavy-meson weak-decay constants $f_{D^*}$ and $f_{B^*}$ are also computed in terms of the heavy-quark spin symmetry. From our calculations, the ratios $f_{D^*}/f_{D}$ and $f_{B^*}/f_{B}$ become only a few $1.01\sim1.03$, which is rather smaller than other theoretical calculations, manifesting the spin symmetry.  
\end{itemize}
%TABLE>>>
As a result, we can conclude that the present phenomenological model, ExNLChQM is quite reliable and promising to investigate the heavy-light quark systems, although  there are several theoretical assumptions and improvements to be addressed from a microscopic point of view. For more detailed tests for the present model, theoretical computations for the decay constants for strange heavy mesons and various structure functions, such as the Isgur-Wise function, are under progress, and the results will appear elsewhere.

%-------------------------------------------------
\section*{acknowledgment}
%-------------------------------------------------
The author is grateful to N.~I.~Kochelev, M.~M.~Musakhanov, H.~-Ch.~Kim, and C.~W.~Kao for fruitful discussions and comments. 
%--------------------------------------------------
%-------------------------------------------------
\section*{appendix}
%-------------------------------------------------
The explicit expression for $\mathcal{U}_5$ in the right-hand-side of Eq.~(\ref{eq:U5}) is given in $(4\times4)$ matrix as follows:
%EQUATION>>>
\begin{equation}
\label{eq:UMAT}
\mathcal{U}_5\approx\left(
\begin{array}{cccc}
1+\frac{i\gamma_5\pi^0}{F_{q\bar{q}}}&
\frac{i\sqrt{2}\gamma_5\pi^+}{F_{q\bar{q}}}&
\frac{i\sqrt{2}\gamma_5\bar{D}^0}{F_{q\bar{Q}}}&
\frac{i\sqrt{2}\gamma_5B^+}{F_{q\bar{Q}}}\\
%--------
\frac{i\sqrt{2}\gamma_5\pi^-}{F_{q\bar{q}}}&
1-\frac{i\gamma_5\pi^0}{F_{q\bar{q}}}&
\frac{i\sqrt{2}\gamma_5D^-}{F_{q\bar{Q}}}&
\frac{i\sqrt{2}\gamma_5B^0}{F_{q\bar{Q}}}\\
%--------
\frac{i\sqrt{2}\gamma_5D^0}{F_{Q\bar{q}}}&
\frac{i\sqrt{2}\gamma_5D^+}{F_{Q\bar{q}}}&
1&0\\
%--------
\frac{i\sqrt{2}\gamma_5B^-}{F_{Q\bar{q}}}&
\frac{i\sqrt{2}\gamma_5\bar{B}^0}{F_{Q\bar{q}}}&
0&1
\end{array}
\right).
\nonumber
\end{equation}
%EQUAITON<<<

The functional integral with two Grassmann variables can be performed as follows:
%EQUATION>>>
\begin{eqnarray}
\int\,D\bar{Q}_vDQ_v Dq D\bar{q}\,e^{i\int d^4x\,\bar{q}\mathcal{A}q+\bar{Q}_v\mathcal{B}Q_v+\bar{Q}_v\mathcal{C}q+\bar{q}\mathcal{C}^\dagger Q_v}
&=&\int\,D\bar{Q}_vDQ_v Dq D\bar{q}\,e^{i\int d^4x\,(\bar{Q}+\bar{q}\mathcal{C}^\dagger\mathcal{B}^\dagger)\mathcal{B}(Q+\mathcal{B}^\dagger\mathcal{C}q)+\bar{q}(\mathcal{A}-\mathcal{C}^\dagger\mathcal{B}^\dagger\mathcal{C})q }
\cr
=\mathrm{Det}[\mathcal{B}]\int\,D\bar{q}Dq\,e^{i\int d^4x\,
\bar{q}(\mathcal{A}-\mathcal{C}^\dagger\mathcal{B}^\dagger\mathcal{C})q }
&=&\mathrm{Sp}\ln[\mathcal{B}]+\mathrm{Sp}\ln[\mathcal{A}-\mathcal{C}^\dagger\mathcal{B}^\dagger\mathcal{C}].
\nonumber
\end{eqnarray}
%EQUAITON<<<
%-------------------------------------------------

%--------------------------------------------------

\begin{thebibliography}{99}
%--------------------------------------------------
\bibitem{Georgi:1990um}
  H.~Georgi,
  %``An Effective Field Theory For Heavy Quarks At Low-energies,''
  Phys.\ Lett.\  {\bf B240}, 447-450 (1990).
%--------------------------------------------------
\bibitem{Neubert:1993mb}
  M.~Neubert,
  %``Heavy quark symmetry,''
  Phys.\ Rept.\  {\bf 245}, 259-396 (1994).
%  [hep-ph/9306320].
%--------------------------------------------------
\bibitem{Beneke:1994sw}
  M.~Beneke, V.~M.~Braun,
  %``Heavy quark effective theory beyond perturbation theory: Renormalons, the pole mass and the residual mass term,''
  Nucl.\ Phys.\  {\bf B426}, 301-343 (1994).
%  [hep-ph/9402364].
%--------------------------------------------------
\bibitem{Abazov:2005ga}
  V.~M.~Abazov {\it et al.} [D0 Collaboration],
  %``Measurement of semileptonic branching fractions of $B$ mesons to narrow $D^{**}$ states,''
  Phys.\ Rev.\ Lett.\  {\bf 95}, 171803 (2005).
%  [hep-ex/0507046].
%--------------------------------------------------
\bibitem{Aubert:2007bx}
  B.~Aubert {\it et al.} [BABAR Collaboration],
  %``A Search for $B^{+} \to \tau^{+} \nu$,''
  Phys.\ Rev.\  {\bf D76}, 052002 (2007).
%  [arXiv:0705.1820 [hep-ex]].
%--------------------------------------------------
\bibitem{Ikado:2006un}
  K.~Ikado {\it et al.} [Belle Collaboration],
  %``Evidence of the Purely Leptonic Decay B- ---> tau- anti-nu(tau),''
  Phys.\ Rev.\ Lett.\  {\bf 97}, 251802 (2006).
%  [hep-ex/0604018].
%--------------------------------------------------
\bibitem{Bai:1999yk}
  J.~Z.~Bai {\it et al.} [BES Collaboration],
  %``Direct measurement of B(D0 ---> phi X0) and B(D+ ---> phi X+),''
  Phys.\ Rev.\  {\bf D62}, 052001 (2000).
%  [hep-ex/9907052].
%--------------------------------------------------
\bibitem{Nakamura:2010zzi}
  K.~Nakamura {\it et al.} [ Particle Data Group Collaboration ],
  %``Review of particle physics,''
  J.\ Phys.\ G {\bf G37}, 075021 (2010).
%--------------------------------------------------
\bibitem{Ebert:1994tv}
  D.~Ebert, T.~Feldmann, R.~Friedrich, H.~Reinhardt,
  %``Effective meson Lagrangian with chiral and heavy quark symmetries from quark flavor dynamics,''
  Nucl.\ Phys.\  {\bf B434}, 619-646 (1995).
%--------------------------------------------------
\bibitem{Cvetic:2004qg}
  G.~Cvetic, C.~S.~Kim, G.~-L.~Wang, W.~Namgung,
  %``Decay constants of heavy meson of 0- state in relativistic Salpeter method,''
  Phys.\ Lett.\  {\bf B596}, 84-89 (2004).
%  [hep-ph/0405112].
%--------------------------------------------------
\bibitem{Wang:2005qx}
  G.~-L.~Wang,
  %``Decay constants of heavy vector mesons in relativistic Bethe-Salpeter method,''
  Phys.\ Lett.\  {\bf B633}, 492-496 (2006).
%  [math-ph/0512009].
%--------------------------------------------------
\bibitem{Ebert:2006hj}
  D.~Ebert, R.~N.~Faustov, V.~O.~Galkin,
  %``Relativistic treatment of the decay constants of light and heavy mesons,''
  Phys.\ Lett.\  {\bf B635}, 93-99 (2006).
%  [hep-ph/0602110].
%--------------------------------------------------
\bibitem{Badalian:2007km}
  A.~M.~Badalian, B.~L.~G.~Bakker, Y.~.A.~Simonov,
  %``Decay constants of the heavy-light mesons from the field correlator method,''
  Phys.\ Rev.\  {\bf D75}, 116001 (2007).
%  [hep-ph/0702157 [HEP-PH]].
%--------------------------------------------------
\bibitem{Hwang:2010hw}
  C.~W.~Hwang,
  %``Analyses of decay constants and light-cone distribution amplitudes for sWave heavy meson,''
  Phys.\ Rev.\  {\bf D81}, 114024 (2010).
%  [arXiv:1003.0972 [hep-ph]].
%--------------------------------------------------
\bibitem{Penin:2001ux}
  A.~A.~Penin, M.~Steinhauser,
  %``Heavy light meson decay constant from QCD sum rules in three loop approximation,''
  Phys.\ Rev.\  {\bf D65}, 054006 (2002).
%  [hep-ph/0108110].
%--------------------------------------------------
\bibitem{Geng:2010df}
  L.~S.~Geng, M.~Altenbuchinger, W.~Weise,
  %``Light quark mass dependence of the $D$ and $D_s$ decay constants,''
  Phys.\ Lett.\  {\bf B696}, 390-395 (2011).
%  [arXiv:1012.0666 [hep-ph]].
%--------------------------------------------------
\bibitem{Allton:1990qg}
  C.~R.~Allton, C.~T.~Sachrajda, V.~Lubicz, L.~Maiani, G.~Martinelli,
  %``A lattice computation of the decay constant of the B meson,''
  Nucl.\ Phys.\  {\bf B349}, 598-616 (1991).
%--------------------------------------------------
\bibitem{Alexandrou:1990dq}
  C.~Alexandrou, F.~Jegerlehner, S.~Gusken, K.~Schilling, R.~Sommer,
  %``B meson properties from lattice QCD,''
  Phys.\ Lett.\  {\bf B256}, 60-67 (1991).
%--------------------------------------------------
\bibitem{Becirevic:1998ua}
  D.~Becirevic {\it et al.}, 
  %P.~Boucaud, J.~P.~Leroy, V.~Lubicz, G.~Martinelli, F.~Mescia, F.~Rapuano,
  %``Nonperturbatively improved heavy - light mesons: Masses and decay constants,''
  Phys.\ Rev.\  {\bf D60}, 074501 (1999).
%  [hep-lat/9811003].
%--------------------------------------------------
\bibitem{Bernard:2009wr}
  C.~Bernard {\it et al.}, 
  %C.~DeTar, M.~Di Pierro, A.~X.~El-Khadra, R.~T.~Evans, E.~D.~Freeland, E.~Gamiz, S.~Gottlieb {\it et al.},
  %``B and D Meson Decay Constants,''
  PoS {\bf LATTICE2008}, 278 (2008).
%  [arXiv:0904.1895 [hep-lat]].
%--------------------------------------------------
\bibitem{AliKhan:2001jg}
  A.~Ali Khan {\it et al.} [CP-PACS Collaboration],
  %``B meson decay constant from two flavor lattice QCD with nonrelativistic heavy quarks,''
  Phys.\ Rev.\  {\bf D64}, 054504 (2001).
%  [hep-lat/0103020].
%--------------------------------------------------
\bibitem{Gray:2005ad}
  A.~Gray {\it et al.} [HPQCD Collaboration],
  %``The B meson decay constant from unquenched lattice QCD,''
  Phys.\ Rev.\ Lett.\  {\bf 95}, 212001 (2005).
%  [hep-lat/0507015].
%--------------------------------------------------
\bibitem{Gamermann:2007fi}
  D.~Gamermann and E.~Oset,
  %``Axial Resonances in the Open and Hidden Charm Sectors,''
  Eur.\ Phys.\ J.\  A {\bf 33}, 119 (2007).
%  [arXiv:0704.2314 [hep-ph]].
%--------------------------------------------------
\bibitem{Staric:2011en}
  M.~Staric {\it et al.} [Belle Collaboration],
  %``Search for CP Violation in D Meson Decays to $\phi \pi^+$,''
  %Submitted to: Phys.Rev.Lett..
  [arXiv:1110.0694 [hep-ex]].
%--------------------------------------------------
\bibitem{Kobayashi:1973fv}
  M.~Kobayashi, T.~Maskawa,
  %``CP Violation in the Renormalizable Theory of Weak Interaction,''
  Prog.\ Theor.\ Phys.\  {\bf 49}, 652-657 (1973).
%--------------------------------------------------
\bibitem{:2008sq}
  B.~I.~Eisenstein {\it et al.} [CLEO Collaboration],
  %``Precision Measurement of B(D+ ---> mu+ nu) and the Pseudoscalar Decay Constant f(D+),''
  Phys.\ Rev.\  {\bf D78}, 052003 (2008).
%  [arXiv:0806.2112 [hep-ex]].
%--------------------------------------------------
\bibitem{Alexander:2009ux}
  J.~P.~Alexander {\it et al.} [CLEO Collaboration],
  %``Measurement of $B{D_s^+ \to \ell^+ \nu}$ and the Decay Constant $fD_s^+$ From 600 $/pb^{-1}$ of $e^\pm$ Annihilation Data Near 4170 MeV,''
  Phys.\ Rev.\  {\bf D79}, 052001 (2009).
% [arXiv:0901.1216 [hep-ex]].
%--------------------------------------------------
\bibitem{Naik:2009tk}
  P.~Naik {\it et al.} [CLEO Collaboration],
  %``Measurement of the Pseudoscalar Decay Constant f(D(s)) Using D(s)+ ---> tau+ nu, tau+ ---> rho+ anti-nu Decays,''
  Phys.\ Rev.\  {\bf D80}, 112004 (2009).
%  [arXiv:0910.3602 [hep-ex]].
%--------------------------------------------------
\bibitem{Rosner:2008yu}
  J.~L.~Rosner, S.~Stone,
  %``Decay Constants of Charged Pseudoscalar Mesons,''
    [arXiv:0802.1043 [hep-ex]].
%--------------------------------------------------
\bibitem{Rosner:2010ak}
  J.~L.~Rosner, S.~Stone,
  %``Leptonic Decays of Charged Pseudoscalar Mesons,''
  [arXiv:1002.1655 [hep-ex]].
%--------------------------------------------------
\bibitem{Nam:2006sx}
  S.~i.~Nam, H.~-Ch.~Kim,
  %``Leading-twist pion and kaon distribution amplitudes in the gauge-invariant nonlocal chiral quark model from the instanton vacuum,''
  Phys.\ Rev.\  {\bf D74}, 076005 (2006).
%  [hep-ph/0609267].
%--------------------------------------------------
\bibitem{Nam:2006mb}
  S.~i.~Nam, H.~-Ch.~Kim,
  %``Twist-3 pion and kaon distribution amplitudes from the instanton vacuum with flavor SU(3) symmetry breaking,''
  Phys.\ Rev.\  {\bf D74}, 096007 (2006).
%  [hep-ph/0608018].
%--------------------------------------------------
\bibitem{Nam:2006au}
  S.~i.~Nam, H.~-Ch.~Kim, A.~Hosaka, M.~M.~Musakhanov,
  %``The Leading-twist pion and kaon distribution amplitudes from the QCD instanton vacuum,''
  Phys.\ Rev.\  {\bf D74}, 014019 (2006).
%  [hep-ph/0605259].
%--------------------------------------------------
\bibitem{Praszalowicz:2001wy}
  M.~Praszalowicz, A.~Rostworowski,
  %``Pion light cone wave function in the nonlocal NJL model,''
  Phys.\ Rev.\  {\bf D64}, 074003 (2001).
%  [hep-ph/0105188].
%--------------------------------------------------
\bibitem{Dorokhov:2010zz}
  A.~Dorokhov,
  %``Photon-pion transition form factor within nonlocal chiral quark model,''
  PoS {\bf LC2010}, 061 (2010).
%--------------------------------------------------
\bibitem{Dorokhov:2010zzb}
  A.~E.~Dorokhov,
  %``Photon-pion transition form factor at high photon virtualities within the nonlocal chiral quark model,''
  JETP Lett.\  {\bf 92}, 707-719 (2010).
%--------------------------------------------------
\bibitem{Dumm:2010hh}
  D.~Gomez Dumm, S.~Noguera, N.~N.~Scoccola,
  %``Pion radiative weak decays in nonlocal chiral quark models,''
  Phys.\ Lett.\  {\bf B698}, 236-242 (2011).
%  [arXiv:1011.6403 [hep-ph]].
%--------------------------------------------------
\bibitem{Noguera:2008cm}
  S.~Noguera, N.~N.~Scoccola,
  %``Nonlocal chiral quark models with wavefunction renormalization: Sigma properties and pi - pi scattering parameters,''
  Phys.\ Rev.\  {\bf D78}, 114002 (2008).
%  [arXiv:0806.0818 [hep-ph]].
%--------------------------------------------------
\bibitem{Shuryak:1981ff}
  E.~V.~Shuryak,
  %``The Role of Instantons in Quantum Chromodynamics. 1. Physical Vacuum,''
  Nucl.\ Phys.\  {\bf B203}, 93 (1982).
%--------------------------------------------------
\bibitem{Diakonov:1983hh}
  D.~Diakonov, V.~Y.~Petrov,
  %``Instanton Based Vacuum from Feynman Variational Principle,''
  Nucl.\ Phys.\  {\bf B245}, 259 (1984).
%--------------------------------------------------
\bibitem{Witten:1978bc}
  E.~Witten,
  %``Instantons, the Quark Model, and the 1/n Expansion,''
  Nucl.\ Phys.\  {\bf B149}, 285 (1979).
%--------------------------------------------------
\bibitem{Diakonov:1995qy}
  D.~Diakonov, M.~V.~Polyakov, C.~Weiss,
  %``Hadronic matrix elements of gluon operators in the instanton vacuum,''
  Nucl.\ Phys.\  {\bf B461}, 539-580 (1996).
%  [hep-ph/9510232].
%--------------------------------------------------
\bibitem{Diakonov:2002fq}
  D.~Diakonov,
  %``Instantons at work,''
  Prog.\ Part.\ Nucl.\ Phys.\  {\bf 51}, 173-222 (2003).
%  [hep-ph/0212026].
%--------------------------------------------------
\bibitem{Goeke:2007bj}
  K.~Goeke, M.~M.~Musakhanov, M.~Siddikov,
  %``Low energy constants of chi PT from the instanton vacuum model,''
  Phys.\ Rev.\  {\bf D76}, 076007 (2007).
%  [arXiv:0707.1997 [hep-ph]].
%--------------------------------------------------
\bibitem{Nam:2010mh}
  S.~i.~Nam, C.~W.~Kao,
  %``Chiral restoration at finite temperature with meson loop corrections,''
  Phys.\ Rev.\  {\bf D82}, 096001 (2010).
%  [arXiv:1005.1689 [hep-ph]].
 %--------------------------------------------------
  \bibitem{Nam:2011vn}
  S.~i.~Nam, C.~W.~Kao,
  %``Chiral restoration at finite T under the magnetic field with the meson-loop corrections,''
  Phys.\ Rev.\  {\bf D83}, 096009 (2011).
%  [arXiv:1103.6057 [hep-ph]].
%--------------------------------------------------
\bibitem{Nam:2008xx}
  S.~i.~Nam, H.~-Ch.~Kim,
  %``Pion weak decay constant at finite density from the instanton vacuum,''
  Phys.\ Lett.\  {\bf B666}, 324-331 (2008).
% [arXiv:0805.0060 [hep-ph]].
%--------------------------------------------------
\bibitem{Dorokhov:2004ze}
  A.~E.~Dorokhov,
  %``Adler function and hadronic contribution to the muon g-2 in a nonlocal
  %chiral quark model,''
  Phys.\ Rev.\  D {\bf 70}, 094011 (2004).
%  [arXiv:hep-ph/0405153].
%--------------------------------------------------
%--------------------------------------------------
%--------------------------------------------------
%--------------------------------------------------
%--------------------------------------------------
%--------------------------------------------------
%--------------------------------------------------
%--------------------------------------------------
%--------------------------------------------------
%--------------------------------------------------
%--------------------------------------------------
%--------------------------------------------------
%--------------------------------------------------
\end{thebibliography}
\end{document}